\documentclass[aps,twocolumn,twoside]{revtex4}

\usepackage{dcolumn}
\usepackage{amsmath}

\usepackage{graphicx}

\def\eref#1{(\ref{#1})}

\def\e{{\rm e}}
\def\O{{\rm O}}
\def\etal{{\it{}et~al.}}
\def\av#1{\langle#1\rangle}

\newlength{\figurewidth}
\begin{document}


\title{Percolation and epidemics in a two-dimensional small world}
\author{M. E. J. Newman}
\affiliation{Santa Fe Institute, 1399 Hyde Park Road, Santa Fe, NM 87501,
U.S.A.}
\author{I. Jensen}
\affiliation{Department of Mathematics and Statistics, University of
Melbourne, Parkville, VIC 3010, Australia}
\author{R. M. Ziff}
\affiliation{Michigan Center for Theoretical Physics and Department
of Chemical Engineering,\\
University of Michigan, Ann Arbor, MI 48109--2136, U.S.A.}
\date{August 31, 2001}

\begin{abstract}
  Percolation on two-dimensional small-world networks has been proposed as
  a model for the spread of plant diseases.  In this paper we give an
  analytic solution of this model using a combination of generating
  function methods and high-order series expansion.  Our solution gives
  accurate predictions for quantities such as the position of the
  percolation threshold and the typical size of disease outbreaks as a
  function of the density of ``shortcuts'' in the small-world network.  Our
  results agree with scaling hypotheses and numerical simulations for the
  same model.
\end{abstract}


\maketitle

\section{Introduction}
\label{intro}
The small-world model has been introduced by Watts and Strogatz~\cite{WS98}
as a simple model of a social network---a network of friendships or
acquaintances between individuals, for instance, or a network of physical
contacts between people through which a disease spreads.  The model
consists of a regular lattice, typically a one-dimensional lattice with
periodic boundary conditions although lattices of two or more dimensions
have been studied as well, with a small number of ``shortcut'' bonds added
between randomly chosen pairs of sites, with density $\phi$ per bond on the
original regular lattice.  The small-world model captures two specific
features observed in real-world networks, namely (1)~logarithmically short
distances through the network between most pairs of individuals and
(2)~high network clustering, meaning that two individuals are much more
likely to be friends with one another if they have one or more other
friends in common.  The model turns out to be amenable to treatment using a
variety of techniques drawn from statistical physics and has as a result
received wide attention in the physics
community~\cite{BA99,NW99b,BW00,MMP00,DM00,KAS00}.

The small-world model has some problems however.  In particular, it is
built on a low-dimensional regular lattice, and there is little
justification to be found in empirical studies of social networks for such
an underlying structure.  As recently pointed out by
Warren~\etal~\cite{WSS01}, however, there is one case in which the
small-world model may be a fairly accurate representation of a real-world
situation, and that is in the spread of plant diseases.  Plant diseases
spread through physical contacts between plants---immediate contagion,
insect vectors, wind, and so forth---and these contacts form a ``social
network'' among the plants in question.  However, plants are sessile, and
confined by and large to the two-dimensional plane of the Earth's surface.
Disease spread as a result of short-range contact between plants is thus
probably well represented as a transmission process on a simple
two-dimensional lattice, and disease spread in which some portion of
transmission is due to longer-range vectors such as wind or insects may be
well represented by a small-world model built upon an underlying
two-dimensional lattice.  Because of this, as well as because of inherent
mathematical interest, a number of recent papers have focused on
two-dimensional small-world networks~\cite{WSS01,WSSKS01,Ozana01}.

In this paper we study bond percolation on two-dimensional small-world
networks.  Bond percolation is equivalent to the standard
susceptible/infectious/recovered (SIR) model of disease
spread~\cite{WSS01,Grassberger83}, in which all individuals are initially
susceptible to the disease, become infected (and hence infectious) with
some probability per unit time if one of their neighbors is infectious, and
recover again, becoming uninfectious and also immune, after a certain time
in the infectious state.  The equivalence of the SIR model to percolation
is straightforward, with the percolation threshold mapping to the epidemic
threshold of the disease in the SIR model, and cluster sizes mapping to the
sizes of disease outbreaks which start with a single disease carrier.
Using a combination of an exact generating function method with a
high-order series expansion, we derive approximate analytic results for the
position of the threshold and the mean outbreak and epidemic sizes as a
function of the density of shortcuts in the 2D small-world network and the
percolation probability, which is equivalent to disease transmission
probability.  As we demonstrate, our results are in excellent agreement
with those from other studies using different methods, as well as with our
own numerical simulations.

\begin{figure}[b]
\begin{center}
\resizebox{\figurewidth}{!}{\includegraphics{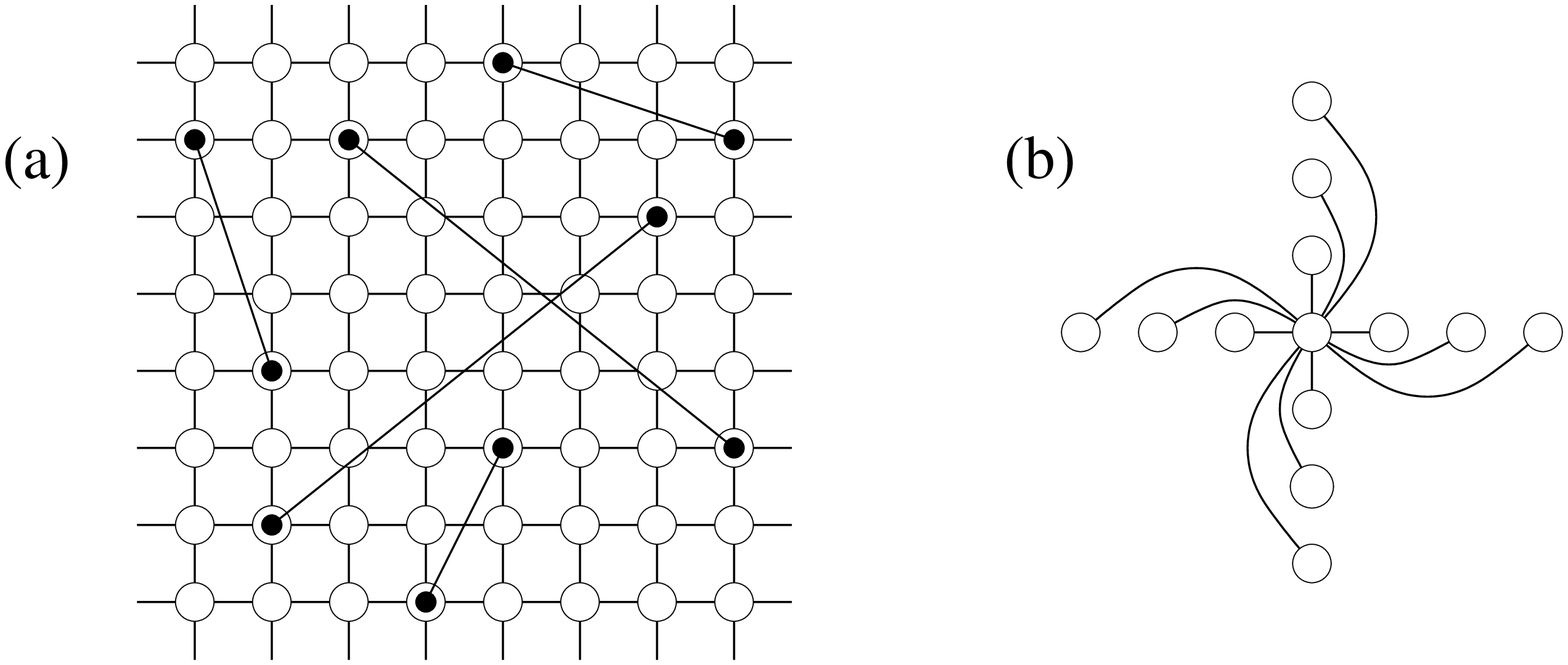}}
\end{center}
\caption{(a) A two-dimensional small-world network built upon a square
  lattice with connection range $k=1$.  (b)~When $k>1$ the underlying
  lattice contains bonds beyond nearest-neighbor bonds along each principal
  axis out to range $k$, as shown here for $k=3$.}
\label{model}
\end{figure}

\section{Generating function formalism}
\label{gf}
We study the two-dimensional small-world model built on the square lattice.
The model is depicted in Fig.~\ref{model}.  We develop our generating
function formalism for the general case of a $d$-dimensional
square/cubic/hypercubic underlying lattice first, narrowing our scope to
the two-dimensional case in Section~\ref{series} where we describe our
series expansion calculations.  For a $d$-dimensional lattice with bonds
along the principal axes out to distance~$k$~\cite{NW99b}, the underlying
lattice has $dL^dk$ bonds on it, where $L$ is the system dimension, and
shortcuts are added with probability $\phi$ per underlying bond, for a
total of $dL^dk\phi$ shortcuts.  Then all bonds, including the shortcut
bonds, are occupied with probability $p$, or not with probability $1-p$,
and we construct the percolation clusters of sites connected by the
occupied bonds.

The generating function part of our calculation follows the method of Moore
and Newman~\cite{MN00}.  We define a probability generating function $H(z)$
thus:
\begin{equation}
H(z) = \sum_{n=1}^\infty P(n) z^n,
\label{hz1}
\end{equation}
where $P(n)$ is the probability that a randomly chosen site in our
small-world network belongs to a connected cluster of $n$ sites other than
the system-spanning cluster.  Note that if the probability distribution
$P(n)$ is properly normalized, then $H(1)=1$ below the transition and
$H(1)=1-S$ above the transition where $S$ is the fraction of the system
occupied by the system-spanning cluster.

We also define $P_0(n)$ to be the probability that a randomly chosen site
belongs to a cluster of $n$ sites on the underlying lattice.  The complete
cluster on the small-world network is composed of a set of such underlying
clusters, joined together by occupied shortcut bonds.  If we denote by
$P(m|n)$ the probability that an underlying cluster of $n$ sites has
exactly $m$ shortcuts emanating from it, then the generating function
$H(z)$ can be written self-consistently as~\cite{MN00}
\begin{equation}
H(z) = \sum_{n=1}^\infty P_0(n) z^n \sum_m P(m|n) [H(z)]^m.
\label{hz2}
\end{equation}
(The derivation of this equation assumes that all clusters other than the
percolating cluster, if there is one, contain no closed loops other than
those on the underlying lattice, i.e.,~there are no loops involving
shortcut bonds.  This is only strictly true in the limit of infinite system
size, and hence our results will only be exact in this limit.)

There are a total of $2dL^dk\phi p$ ends of occupied shortcuts in the
model, and since all ends are uniformly distributed over the lattice the
probability that any end lands in a given cluster of size $n$ is just
$n/L^d$ for large $L$.  Thus $P(m|n)$ is given by the binomial distribution
\begin{equation}
P(m|n) = {2dL^dk\phi p\choose m} \biggl[{n\over L^d}\biggr]^m
         \biggl[1-{n\over L^d}\biggr]^{2dL^dk\phi p-m}.
\end{equation}
Substituting into Eq.~\eref{hz2} and performing the sum over $m$, this
gives
\begin{eqnarray}
H(z) &=& \sum_n P_0(n) z^n \biggl[ 1 + (H(z)-1){n\over
         L^d}\biggr]^{2dL^dk\phi p}\nonumber\\
     &=& \sum_n P_0(n) \bigl[z \e^{2dk\phi p(H(z)-1)}\bigr]^n,
\label{hz3}
\end{eqnarray}
where the last equality holds in the limit of large~$L$.  If we define the
additional generating function
\begin{equation}
H_0(z) = \sum_n P_0(n) z^n,
\end{equation}
which is the probability generating function for the sizes of clusters for
ordinary bond percolation on the underlying lattice, then Eq.~\eref{hz3}
can be written in the form
\begin{equation}
H(z) = H_0\bigl(z\e^{2dk\phi p(H(z)-1)}\bigr).
\label{hzfinal}
\end{equation}
This gives a self-consistency condition from which we can evaluate $H(z)$,
and hence we can evaluate the probabilities $P(n)$ for cluster sizes in the
small-world model.

In fact, it is rarely possible to solve Eq.~\eref{hzfinal} for $H(z)$ in
closed form (although evaluation by numerical iteration is often feasible),
but we can derive closed-form expressions for other quantities of interest.
In particular, the average size of the cluster to which a randomly chosen
site belongs is given by
\begin{equation}
\av{n} = \sum_n n P(n) = H'(1) = H_0'(1) [1 + 2dk\phi p H'(1)],
\end{equation}
or, rearranging,
\begin{equation}
\av{n} = {H_0'(1)\over1-2dk\phi p H_0'(1)}.
\label{avs}
\end{equation}
This quantity diverges when
\begin{equation}
2dk\phi p H_0'(1) = 1,
\end{equation}
or equivalently when
\begin{equation}
\phi = {1\over2dk p H_0'(1)},
\label{phic}
\end{equation}
and this point marks the phase transition at which a giant cluster scaling
as the size of the entire system first forms.  Another way of looking at
this result is to note that $H_0'(1)=\av{n_0}$, the average cluster size on
the underlying lattice.  Thus percolation takes place when
\begin{equation}
2dk\phi p = {1\over\av{n_0}}.
\label{rg}
\end{equation}
The quantity on the left-hand side of this equation is the average density
of the ends of occupied shortcut bonds on the lattice (see
Section~\ref{scaling}), and thus percolation takes place when there is
exactly one end of an occupied shortcut bond per cluster on the underlying
lattice.  This is reminiscent of the phase transition in an
Erd\H{o}s--R\'enyi random graph, which occurs at the point where each
vertex in the graph is attached to exactly one edge~\cite{Bollobas85}.

Above the phase transition $S=1-H(1)$ is the size of the giant cluster.
Setting $z=1$ in Eq.~\eref{hzfinal} we find that $S$ is a solution of
\begin{equation}
S = 1 - H_0\bigl(\e^{-2dk\phi p S}\bigr),
\label{giantcomp}
\end{equation}
which can be solved by numerical iteration starting from a suitable initial
value of~$S$.  An expression similar to~\eref{avs} for the average size of
non-percolating clusters above the transition can also be derived.  See,
for example, Ref.~\onlinecite{NSW01}.

\section{Series expansions}
\label{series}
In the one-dimensional case studied in Ref.~\onlinecite{MN00}, the
calculation of the generating function $H_0(z)$ is trivial---it is
equivalent to solving the problem of bond percolation in one dimension.  In
the present case however we are interested primarily in the two-dimensional
small-world model, and calculating $H_0$ is much harder; no exact solution
has ever been given for the distribution of cluster sizes for bond
percolation on the square lattice.  Instead, therefore, we turn to series
expansion to calculate $H_0$ approximately.

$H_0(z)$ for the two-dimensional case can be written as
\begin{equation}
H_0(z) = \sum_{stn} n z^n p^s (1-p)^t g_{stn},
\end{equation}
where $g_{stn}$ is the number of different possible clusters on a square
lattice which have $s$ occupied bonds, $t$ unoccupied bonds around their
perimeter, and $n$ sites.  If we can calculate $g_{stn}$ up to some finite
order then we can calculate an approximation to $H_0(z)$ also.
Unfortunately, because $g_{stn}$ depends separately on three different
indices, it is prohibitively memory-intensive to calculate on a computer up
to high order.  We note however that we only need $H_0$ as a function of
$p$ and $z$, and not of $1-p$ separately, so we can collect terms in $p$
and rewrite the generating function as
\begin{equation}
H_0(z) = \sum_{m=0}^\infty p^m Q_m(z),
\end{equation}
where the quantities $Q_m(z)$ are finite polynomials in $z$ which are, it's
not hard to show, of order $z^{m+1}$.  Calculating these polynomials is
considerably more economical than calculating the entire set of $g_{stn}$.
We have calculated them up to order $m=31$ using the finite-lattice
method~\cite{Enting96}, in which a generating function for the infinite
lattice is built up by combining generating functions for the same problem
on finite lattices.  The finite lattices used in this case were rectangles
of $h\times l$ sites and the quantity we consider is the fundamental
generating function for the cluster density
\begin{equation}
G(z,p) = \sum_{stn} z^n p^s (1-p)^t g_{stn},
\end{equation}
which, it can be shown, is given by the linear combination
\begin{equation}
G(z,p) = \sum_{hl} w_{hl} G_{hl}(z,p),
\end{equation}
where $w_{hl}$ are constant weights that are independent of both $z$ and
$p$, and $G_{hl}(z,p)$ is the generating function for connected clusters
(bond animals) which span an $h\times l$ rectangle both from left to right
and from top to bottom.

Due to the symmetry of the square lattice, the weight factor $w_{hl}$ is
simply
\begin{equation}
w_{hl} = \left\lbrace\begin{array}{ll}
           0 & \qquad\mbox{for $l<h$} \\
           1 & \qquad\mbox{for $l=h$} \\
           2 & \qquad\mbox{for $l>h$.}
         \end{array}\right.
\end{equation}
The individual generating functions $G_{hl}(z,p)$ for the finite lattices
are calculated using a transfer matrix method with generating functions for
all rectangles of a given height $h$ being evaluated in a single
calculation.  The algorithm we use is based on that of
Conway~\cite{Conway95} with enhancements similar to those used by
Jensen~\cite{Jensen01} for the enumeration of site animals on the square
lattice~\cite{note1}.  Since clusters spanning a rectangle of $h\times l$
sites contain at least $h+l-2$ bonds, we must calculate $G_{hl}(z,p)$ for
all $h\leq (m+1)/2$ and $h\leq l\leq m-h+2$ in order to derive a series
expansion for $G(z,p)$ correct to order $m$ in~$p$.  For the order $m=31$
calculation described here the maximal value of $h$ required was~$16$.

\begin{table*}[t]
\setlength{\tabcolsep}{4pt}
\begin{center}
\begin{tabular}{r|l}
 $m$ & $Q_m(z)$ \\
\hline
  0  & $z$ \\
  1  & $4 z^2 - 4 z$ \\
  2  & $18 z^3 - 24 z^2 + 6 z$ \\
  3  & $88 z^4 - 144 z^3 + 60 z^2 - 4 z$ \\
  4  & $435 z^5 - 860 z^4 + 504 z^3 - 80 z^2 + z$ \\
  5  & $2184 z^6 - 5020 z^5 + 3784 z^4 - 1008 z^3 + 60 z^2$ \\
  6  & $11018 z^7 - 28932 z^6 + 26550 z^5 - 9872 z^4 + 1260 z^3 - 24 z^2$ \\
  7  & $55888 z^8 - 164668 z^7 + 177972 z^6 - 85100 z^5 + 16912 z^4 - 1008 z^3 + 4 z^2$ \\
  8  & $284229 z^9 - 928840 z^8 + 1153698 z^7 - 673836 z^6 + 184125 z^5 - 19880 z^4 + 504 z^3$ \\
  9  & $1448800 z^{10} - 5197176 z^9 + 7291488 z^8 - 5030312 z^7 + 1754424 z^6 - 283320 z^5 + 16240 z^4 - 144 z^3$ \\
 10  & $7396290 z^{11} - 28890160 z^{10} + 45155952 z^9 - 35926720 z^8 + 15278872 z^7 - 3323088 z^6 + 317940 z^5 - 9104 z^4 + 18 z^3$ \\
\end{tabular}
\end{center}
\caption{The values of the polynomials $Q_m(z)$ up to $m=10$.}
\label{qmtab}
\end{table*}

Once $G(z,p)$ is calculated, the polynomials $Q_m(z)$ are easily extracted
by collecting terms in~$p$.  (Alternatively, one could write
$H_0(z)=z\,\partial G/\partial z$, although doing so offers no operational
advantage in the present case.)  In Table~\ref{qmtab} we list the values of
$Q_m(z)$ for $m$ up to~10; the complete set of polynomials up to $m=31$ is
available from the authors on request.  We notice that since $H_0(1)=1$ for
all $p<p_c$, as it must given that the probability distribution it
generates is properly normalized, it must be the case that
\begin{equation}
Q_m(1) = \biggl\lbrace \begin{array}{ll}
           1        & \mbox{for $m=0$}\\
           0 \qquad & \mbox{for $m\ge1$.}
         \end{array}
\end{equation}
It can easily be verified that this is true for the orders given in
Table~\ref{qmtab}.  This implies that $H_0$ will be correctly normalized
even if we truncate its series at finite order in~$p$, as we do here.  This
makes our calculations a little easier.

In order to calculate the average cluster size~\eref{avs} and position of
the phase transition~\eref{phic} in our small-world model, we need to
evaluate the quantity $H_0'(1)$, which is given by
\begin{equation}
H_0'(1) = \sum_{m=0}^\infty p^m Q_m'(1).
\label{h0series}
\end{equation}
The quantities $Q_m'(1)$ are just numbers---their values up to $m=31$ are
given in Table~\ref{qmptab}---so that this expression is a simple power
series in~$p$.  If we make use of our results for $m$ up to 31 to evaluate
this quantity directly, we can calculate the behavior of the 2D small-world
model using the results of Section~\ref{gf}.  However we can do better than
this.

\begin{table}[b]
\setlength{\tabcolsep}{4pt}
\begin{center}
\begin{tabular}{rr|rr}
 $m$ & $Q_m'(1)$ & $m$ & $Q_m'(1)$ \\
\hline
   0 &             $1$ &   1 &             $4$ \\
   2 &            $12$ &   3 &            $36$ \\
   4 &            $88$ &   5 &           $236$ \\
   6 &           $528$ &   7 &          $1392$ \\
   8 &          $2828$ &   9 &          $7608$ \\
  10 &         $14312$ &  11 &         $39348$ \\
  12 &         $69704$ &  13 &        $197620$ \\
  14 &        $318232$ &  15 &       $1013424$ \\
  16 &       $1278912$ &  17 &       $5362680$ \\
  18 &       $4418884$ &  19 &      $28221636$ \\
  20 &      $11543548$ &  21 &     $152533600$ \\
  22 &     $-20880672$ &  23 &     $903135760$ \\
  24 &    $-705437704$ &  25 &    $5680639336$ \\
  26 &   $-7577181144$ &  27 &   $37205966052$ \\
  28 &  $-66485042424$ &  29 &  $253460708032$ \\
  30 & $-534464876516$ &  31 & $1767651092388$ \\
\end{tabular}
\end{center}
\caption{The derivatives $Q_m'(1)$ for all orders up to $m=31$, which are
  also the coefficients of the series for the mean cluster size---see
  Eq.~\eref{h0series}.  Note that although the values for $Q_m'(1)$ appear
  initially to be positive and increasing roughly exponentially, this rule
  does not hold in general.  The first negative value is at $m=22$, and
  the signs of $Q_m'(1)$ appear to alternate for $m>22$.}
\label{qmptab}
\end{table}

\begin{figure}[t]
\begin{center}
\resizebox{\figurewidth}{!}{\includegraphics{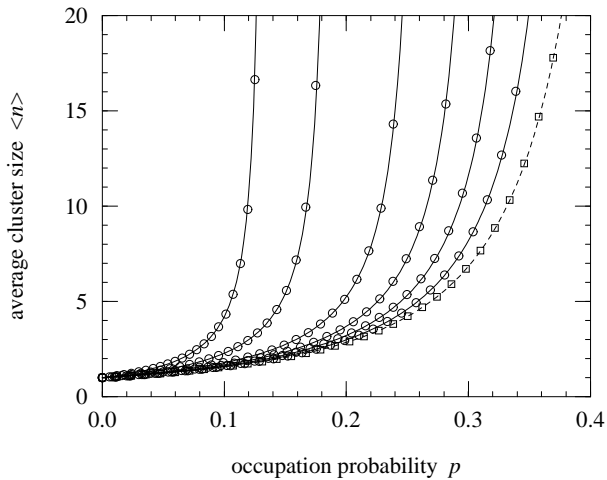}}
\end{center}
\caption{The average of size in sites of the cluster to which a randomly
  chosen site belongs for bond percolation on a two-dimensional small-world
  network with $k=1$.  The circles are simulation results for systems of
  $1024\times1024$ sites, calculated using the fast algorithm of Newman and
  Ziff~\cite{NZ00}, and the solid lines are the analytic result,
  Eq.~\eref{avs}.  From left to right, the values of $\phi$ for each of the
  lines are 1.0, 0.5, 0.2, 0.1, 0.05, and~0.02.  As $\phi$ diminishes the
  lines asymptote to the normal square lattice form which is indicated by
  the square symbols (simulation results) and the dotted line (series
  expansion/Pad\'e approximant result).}
\label{avsfig}
\end{figure}

Since $H_0'(1)$ is the average size $\av{n_0}$ of a cluster in ordinary
bond percolation on the square lattice, we know that it must diverge at
$p_c=\frac12$, and that it does so as $(p_c-p)^{-\gamma}$, where $\gamma$
is the mean cluster-size exponent for two-dimensional percolation which is
equal to $\frac{43}{18}$.  With this information we can construct a Pad\'e
approximant to $H_0'(1)$~\cite{EF63,GG74}.  Writing
\begin{equation}
H_0'(1) = A(p) \biggl[{p_c-p\over p_c}\biggr]^{-\gamma},
\label{pade1}
\end{equation}
where $A(p)$ is assumed analytic near $p_c$, we construct a Pad\'e
approximant to the series for
\begin{equation}
A(p) = \biggl[{p_c-p\over p_c}\biggr]^\gamma H_0'(1),
\label{pade2}
\end{equation}
using our series for $H_0'(1)$.  Then we use this approximant in
Eq.~\eref{pade1} to give an expression for $H_0'(1)$ which agrees with our
series expansion result to all available orders, and has a divergence of
the expected kind at $p=\frac12$.  As is typically the case with Pad\'e
approximants, the best approximations are achieved with the highest order
symmetric or near-symmetric approximants and using all available orders in
our series expansion, we find the best results using a $[15,15]$
approximant to $A(p)$ in Eq.~\eref{pade2}.  In Fig.~\ref{avsfig} we show
the resulting estimate for $\av{n_0}=H_0'(1)$ (dotted line) as a function
of $p$ against numerical results for the average cluster size on an
ordinary square lattice (squares).  As the figure shows, the agreement is
excellent.

\begin{figure}[t]
\begin{center}
\resizebox{\figurewidth}{!}{\includegraphics{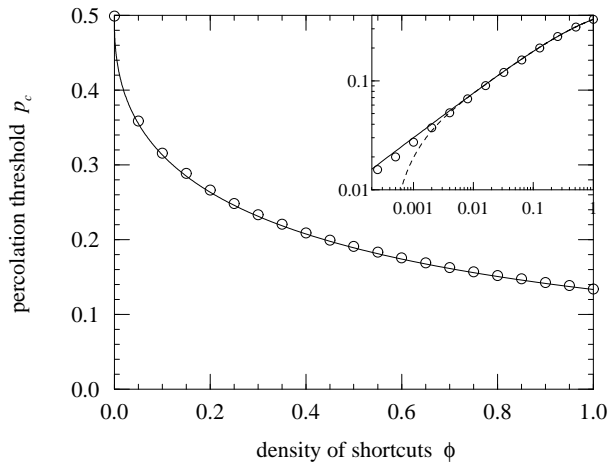}}
\end{center}
\caption{The position of the percolation transition for the 2D small-world
  network as a function of the density $\phi$ of shortcuts.  The points are
  simulation results, again using the algorithm of Ref.~\onlinecite{NZ00},
  and the lines are our analytic calculation using a Pad\'e approximant.
  For the simulations, the value of $p_c$ was taken to be the point at
  which the size of the largest cluster in the system has maximal gradient
  as a function of~$p$.  The slight difference between the numerical and
  analytic results appears to be a systematic error in the estimation of
  $p_c$ from the numerical results (see Ref.~\onlinecite{NW99b} for a
  discussion of this point).  Inset: the same comparison on logarithmic
  scales.  In this case, the vertical axis measures $\frac12-p_c$, which
  should go to zero with $\phi$ according to
  $\frac12-p_c\sim\phi^{1/\gamma}$ where $\gamma=\frac{43}{18}$.  (This can
  be deduced from Eq.~\eref{phic}, and is also shown in
  Ref.~\onlinecite{WSS01}.)  Again the solid line is the Pad\'e approximant
  calculation, while the dotted line represents the value of $p_c$
  calculated directly from the series expansion without using a Pad\'e
  approximant.}
\label{pc}
\end{figure}

Substituting our Pad\'e approximant expression for $H_0'(1)$ into
Eq.~\eref{avs} we can now calculate average cluster size for the
small-world model for any value of $\phi$, and from Eq.~\eref{phic} we can
calculate the position of the phase transition.  In Fig.~\ref{avsfig} we
show the results for $\av{n}$ as a function of $p$ along with numerical
results for the same quantity from simulations of the model.  In
Fig.~\ref{pc} we show the results for $p_c$ plotted against numerical
calculations~\cite{note2}.  In both cases, the agreement between analytic
and numerical results is excellent.  In the inset of Fig.~\ref{pc} we show
the results for $p_c$ on logarithmic scales, along with the value
calculated by using the series expansion for $H_0'(1)$ directly in
Eq.~\eref{phic} (dotted line).  As the figure shows, the Pad\'e approximate
continues to be accurate to very low values of $\phi$, where the direct
series expansion fails.

\section{Scaling forms}
\label{scaling}
As Ozana~\cite{Ozana01} has pointed out, there are two competing
length-scales present in percolation models on small-world networks.  One
is the characteristic length $\xi$ of the small-world model itself which is
given by $\xi=1/(2\phi k d)^{1/d}$, where $d$ is the dimension of the
underlying lattice (2~in the present case).  This length is the typical
linear dimension of the volume on the underlying lattice which contains the
end of one shortcut on average.  In other words $\xi^{-d} = 2\phi k d$ is
the density of the ends of shortcuts on the lattice.  (In fact, one
normally leaves the factor of 2 out of the definition of the characteristic
length, but we include it since it makes the resulting formulas somewhat
neater in our case.)  In the current percolation model, only a fraction $p$
of the shortcuts are occupied and thereby contribute to the behavior of the
model---the unoccupied shortcuts can be ignored.  Thus the appropriate
characteristic length in our case is derived from the density of ends of
occupied shortcuts, which was discussed in Section~\ref{gf}.  The correct
expression is
\begin{equation}
\xi = {1\over(2p\phi kd)^{1/d}}.
\label{defsxi}
\end{equation}

The other length scale is the correlation length or typical cluster
dimension for the percolation clusters on the underlying lattice.  Normally
the latter is also denoted $\xi$, but to avoid confusion we follow the
notation of Ref.~\onlinecite{Ozana01} and here denote it~$\zeta$.  Ozana
has given a finite-size scaling theory for percolation on small-world
networks which addresses the interaction of each of these length-scales
with the lattice dimension~$L$.  Our analytic calculations however treat
the case of $L\to\infty$, for which a simpler scaling theory applies.  In
this case, the only dimensionless combination of lengths is the ratio
$\zeta/\xi$, and any observable quantity $\mathcal{Q}$ must satisfy a
scaling relation of the form
\begin{equation}
\mathcal{Q} \sim \zeta^{d\alpha} \xi^{-d\beta} f(\zeta/\xi).
\label{scaling1}
\end{equation}
where $\alpha$ and $\beta$ are scaling exponents and $f(x)$ is a universal
scaling function.  This form applies when we are in the region where both
$\xi$ and $\zeta$ are much greater than the lattice constant, i.e.,~when
shortcut density is low (the scaling region of the small-world model) and
when we are close to the percolation transition on the normal square
lattice.

We can rewrite Eq.~\eref{scaling1} in a simpler form by making use of
Eq.~\eref{defsxi} and the fact that the typical cluster dimension $\zeta$
on the underlying lattice is related to typical cluster volume by
$\zeta^d=\av{n_0}$ (assuming compact clusters).  This then implies that
\begin{equation}
\mathcal{Q} \sim \av{n_0}^\alpha (p\phi kd)^\beta F(2 p\phi kd\av{n_0}),
\label{scaling2}
\end{equation}
where $F(x)$ is another universal scaling function.

Consider for example the average cluster size $\av{n}$ for the small-world
model.  Since we know this becomes equal to its normal square-lattice value
$\av{n_0}$ when $\phi=0$, we can immediately assume $\alpha=1$, $\beta=0$
and
\begin{equation}
{\av{n}\over\av{n_0}} = F(2p\phi kd\av{n_0}).
\label{scaling3}
\end{equation}
Thus, a plot of $\av{n}/\av{n_0}$ against the scaling variable
$x\equiv2p\phi kd\av{n_0}$ should yield a data collapse whose form follows
the scaling function~$F(x)$.  In fact there is no need to make a scaling
plot in this simple case.  Comparison of Eq.~\eref{scaling3} with
Eq.~\eref{avs}, bearing in mind that $\av{n_0}=H_0'(1)$, reveals that
$\av{n}$ does indeed follow the expected scaling form with
\begin{equation}
F(x) = {1\over1-x}.
\end{equation}
The point $x=1$, which is also the point at which the two length-scales are
equal $\xi=\zeta$, thus represents the percolation transition in this
case.  (This observation is equivalent to Eq.~\eref{rg}.)

A slightly less trivial example of the scaling form~\eref{scaling2} is the
scaling of the size $S$ of the giant percolation cluster,
Eq.~\eref{giantcomp}.  To deduce the leading terms in the scaling relation
for~$S$, we expand~\eref{giantcomp} close to the percolation transition in
powers of~$S$, to give
\begin{equation}
S = xS - {[H_0'(1)+H_0''(1)]\over2[H_0'(1)]^2}\, x^2 S^2 + \O(S^3).
\end{equation}
Rearranging and keeping terms to leading order, we find
\begin{equation}
S \simeq {2[H_0'(1)]^2\over[H_0'(1)+H_0''(1)]} \left[{x-1\over x^2}\right]
  = 2 {\av{n_0}^2\over\av{n^2_0}} \left[{x-1\over x^2}\right].
\end{equation}
If there is only one correlation length for percolation on the underlying
lattice then $\av{n_0}^2/\av{n^2_0}$ is homogeneous in it and hence
constant in the critical region.  Thus $S$ scales as $S\sim(x-1)/x^2$ close
to the transition, with the leading constant being zero below the
transition and $\O(1)$ above it.

\section{Conclusions}
We have presented analytic results for bond percolation in the
two-dimensional small-world network model, which has been proposed as a
simple model of the spread of plant diseases.  Using a combination of
generating function methods and series expansion, we have derived
approximate but highly accurate expressions for quantities such as the
position of the percolation transition in the model, the typical size of
non-percolating clusters, and the typical size of the percolating cluster.
Our results are in excellent agreement with numerical simulations of the
model.  By judicious use of Pad\'e approximants for the series expansions,
the results can even be extended to very low shortcut densities, where a
simple series expansion fails.  The results are also in good agreement with
the expected scaling forms for the model.

\begin{acknowledgments}
  The authors would like to thank Len Sander for helpful conversations and
  Marek Ozana for providing an early preprint of Ref.~\onlinecite{Ozana01}.
  This work was funded in part by the US National Science Foundation and
  the Australian Research Council.  Thanks are also due to the Australian
  Partnership for Advanced Computing for their generous allocation of
  computing resources.
\end{acknowledgments}


\begin{thebibliography}{9}
\bibitem{note1}
  Two practical differences between the algorithm we use and the algorithms
  used in previous calculations are that we use a transfer matrix with an
  intersection which cuts through the edges of the lattice rather than the
  vertices and we calculate a two parameter ($z$~and~$p$) generating
  function rather than just a one parameter~($p$) generating function.
  Most other aspects such as the encoding and transformations of the
  configurations of occupied and empty edges along the intersection and the
  rules for updating the associated generating functions are similar to
  earlier work.
\bibitem{note2}
  In fact, the solid curves for $\av{n}$ in Fig.~\ref{avsfig} change little
  if one uses the series expansion, Eq.~\eref{h0series}, directly in
  Eq.~\eref{avs}, rather than using the Pad\'e approximant.  As the inset
  of Fig.~\ref{pc} shows, the difference between the series and the Pad\'e
  approximant only becomes important for very small values of the shortcut
  density $\phi\lesssim0.002$, and the curves in Fig.~\ref{avsfig} do not
  reach such small values (except for the dotted $\phi=0$ curve).
\end{thebibliography}
\end{document}